\documentclass[Times,4pt,aps,pra,twocolumn,showpacs,amsmath,amssymb,floatfix,footinbib,superscriptaddress]{revtex4-2}
\usepackage{amsmath,natbib,graphicx,subfigure,amssymb,graphics,amsmath,mathrsfs,CJK,color,comment}
\usepackage{multirow,fancyhdr,color,bm,tabularx,psfrag,geometry,dcolumn,datetime,physics,booktabs}
\usepackage[colorlinks=true,linkcolor=blue,urlcolor=blue,citecolor=blue]{hyperref}
\usepackage[mathlines]{lineno}
\def\footnoterule{\kern -1mm \hrule width 5.8cm \kern 2.2mm}
\usepackage{tikz,xcolor,hyperref}% Make Orcid icon
\definecolor{lime}{HTML}{A6CE39}
\DeclareRobustCommand{\orcidicon}{%
    \begin{tikzpicture}
    \draw[lime, fill=lime] (0,0)
    circle [radius=0.16]
    node[white] {{\fontfamily{qag}\selectfont \tiny ID}};\draw[white, fill=white] (-0.0625,0.095)
    circle [radius=0.007];
    \end{tikzpicture}
    \hspace{-2mm}}
\foreach \x in {A, ..., Z}
{\expandafter\xdef\csname orcid\x\endcsname{\noexpand\href{https://orcid.org/\csname orcidauthor\x\endcsname}{\noexpand\orcidicon}}}

\begin{document}

%\footnotesize，\scriptsize，\ tiny 等，大小依次递减
\title{ Adjusting the left-handedness in a cold \(^{87}\)Rb atom via multiple parameter modulation\footnote{ Supported
by the National Natural Science Foundation of China ( Grant Nos. 61205205 and 6156508508 ), the General Program of
Yunnan Provincial Research Foundation of Basic Research for application, China ( Grant No. 2016FB009 ) and the
Foundation for Personnel training projects of Yunnan Province, China ( Grant No. KKSY201207068 ).}}

%\thanks{}

\author{ShunCai Zhao\orcidA{}}
\email[Corresponding author: ]{zscgroup@kmust.edu.cn, zscnum1@126.com.}
\affiliation{Physics department, Kunming University of Science and Technology, Kunming, 650500, PR China}
\affiliation{Center for Quantum Materials and Computational Condensed Matter Physics, Faculty of Science, Kunming University of Science and Technology, Kunming, 650500, PR China}
\author{Qi-Xuan Wu}
\affiliation{College English department, Kunming University of Science and Technology, Kunming, 650500, PR China}
\author{Kun Ma}
\affiliation{Physics department, Kunming University of Science and Technology, Kunming, 650500, PR China}
\affiliation{Center for Quantum Materials and Computational Condensed Matter Physics, Faculty of Science, Kunming University of Science and Technology, Kunming, 650500, PR China}

\begin{abstract}
We demonstrate the adjusting left-handedness in the cold \(^{87}\)Rb atom by its number density, the strong coupling field and
two incoherent pumping fields. The results show that more dense \(^{87}\)Rb atoms and more stronger coupling field
can influence the left-handedness more greatly, while the increasing two incoherent pumping fields construct the negative magnetic response but
depress the negative electric response. The left-handedness adjusted by multiple parameter in the cold \(^{87}\)Rb atomic system provides the flexibility and feasibility for the coming experiment.
\begin{description}
%\item[PACS: ]{78.20.Ci, 42.50.Gy, 81.05.Xj, 78.67.Pt }
\item[Keywords:]{left-handedness; cold $^{87}$Rb atom;  multiple parameter modulation}
\end{description}
\end{abstract}
%%%%%%%%%%%%%%%%%%%%%%%%%%%%%%%%%%%%%%%%%%%%%%%%%%%%%%%%%%%%%%%%%%%%%%%%%%%%%%%%%%%%%%%%%%%%%%%%%%%%%%%%%%%%%%%%%%%%%%%%%%%%%%
%\date{\currenttime,~\today}
\maketitle
%\tableofcontents

\section{Introduction}
\label{Introduction}
\linespread{0.5}
Because of its complicated hyperfine structures associated with $^{87}$Rb atom, the cold \(^{87}\)Rb atom plays a key role in quantum optics and quantum communication\cite{1,2,3,4,5,6,7} in the past few decades. Great progress has been made in the application research of cold $^{87}$Rb atom from both theoretical and experimental viewpoints. For some selected transitions in cold \(^{87}\)Rb atom, several recent theoretical and experimental studies were carried out about Electromagnetically induced transparency(EIT)\cite{8,9,10,11,12} and Bose-Einstein condensation(BEC)\cite{13,14,15}, and so on.

In this work, via the selecting appropriately hyperfine transitions of the $^{87}$Rb atom\cite{16}, we exploit left-handedness\cite{17} in a four-level atomic system realized in cold \(^{87}\)Rb atom, i.e., over some certain frequency ranges the four-level $^{87}$Rb atomic system exhibits simultaneously negative relative permittivity \(\varepsilon_r\) and permeability \(\mu_r\).  And the left-handedness can be adjusted by the different pumping frequencies of the two incoherent pumping fields under the different number density and different strength of the coupling field. It is found that the incoherent pumping fields play very important roles in realizing left-handedness, which was also proved to realized gain-assisted negative group-velocity index in recent years\cite{18}. The left-handedness achieved in the cold $^{87}$Rb atomic system by the simulative four-level atomic system increases the feasibility in the coming experimental realization.

Our paper is organized as follows: Section 2 gives the simulative model for $^{87}$Rb hyperfine structure with some experimental parameters\cite{10}. Section 3 we provide the numerical results and analysis to the possible experimental realization cold$^{87}$Rb atoms in our scheme. Finally, Section 4 provides the conclusions.

\section{The simulating model for $^{87}$Rb hyperfine structure }

\begin{figure}[htp]
\centering
\includegraphics[width=3.0in]{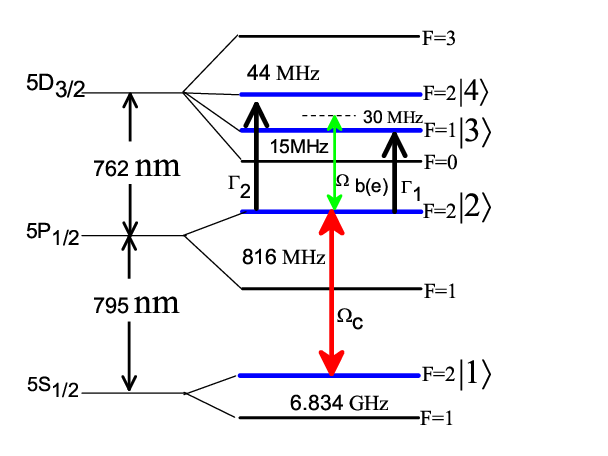}
\caption{(Color online) The levels $|5S_{1/2},F=2\rangle$, $|5P_{1/2},F=2\rangle$, $|5D_{3/2},F=1\rangle$, $|5D_{3/2},F=2\rangle$ in $^{87}$Rb atomic hyperfine
structure were labeled as $|1\rangle$, $|2\rangle$, $|3\rangle$ and  $|4\rangle$ in a four-level atomic system. Two incoherent pumping fields $\Gamma_{1}$, $\Gamma_{2}$ and the electric and magnetic components of the probe field with $\Omega_{e}$ and $\Omega_{b}$ drive the transitions $|2\rangle$-$|3\rangle$ and $|2\rangle$-$|4\rangle$ simultaneously.The ground level $|1\rangle$ and the intermediate level $|2\rangle$ are coupled by a strong coupling field with the Rabi frequency $\Omega_{c}$.}
\label{1}
\end{figure}

In the proposed simulative model, we should point out that such a four-level atomic configuration can be realized in cold $^{87}$Rb atoms(nuclear spin I=3/2) using the 5S-5P-5D hyperfine structure\cite{16}. Figure 1 shows the $^{87}$Rb atomic hyperfine structure, in which the energy levels $|5S_{1/2}, F=2\rangle$, $|5P_{1/2}, F=2\rangle$, $|5D_{3/2}, F=1\rangle$, $|5D_{3/2}, F=2\rangle$ were labeled as $|1\rangle$, $|2\rangle$, $|3\rangle$ and  $|4\rangle$, respectively. And the four-level atomic system has one stable ground state $|1\rangle$, an intermediate level $|2\rangle$ and two excited states $|3\rangle$ and $|4\rangle$. The ground level $|1\rangle$ and the intermediate level $|2\rangle$ are coupled by a strong coupling field $\omega_{c}$ with the Rabi frequency $\Omega_{c}$. The magnetic component of a left-circular polarization probe field ($\Omega_{b}$ )couples to transition $|2\rangle$-$|4\rangle$ with Rabi frequency $\Omega_{b}$ = $\vec{\mu}_{24} \cdot \vec{B}/\hbar$, where $\vec{\mu}_{24}$ is the corresponding magnetic dipole moment. Because of the parity selection rules, the electric component of this probe field ( $\Omega_{e}$ ) couples to transition $|2\rangle$-$|3\rangle$ with Rabi frequency $\Omega_{e}$ = $\vec{d}_{32} \cdot \vec{E}/\hbar$, where $\vec{d}_{32}$ is the electric dipole moment.
Simultaneously, two incoherent pumping fields ( $\Gamma_{1}$ and  $\Gamma_{2}$ ) which can be provided by the diode laser with a broad variable line-width are applied to the transitions $|2\rangle$-$|3\rangle$ and $|2\rangle$-$|4\rangle$, respectively. It should be noted that the polarizations of the two incoherent pumping fields $\vec{\varepsilon}_{1}$ and $\vec{\varepsilon}_{2}$ were properly arrange in such a way that $\vec{\varepsilon}_{1} \cdot \vec{d}_{32}$= $\vec{\varepsilon}_{2} \cdot \vec{d}_{42}$=0, i.e., one field acts on only one transition. According to this, the interference terms induced by the incoherent pumping fields are not included in the density matrix equations\cite{19}.

Under the electric dipole and rotating-wave approximations the density-matrix equations of motion for this system can be written as  $\frac{d\rho}{dt}=-\frac{i}{\hbar}[H,\rho]+\Lambda\rho $ in Eq.(1), where $\Lambda\rho$ represents the irreversible decay part of the four-level $^{87}$Rb atomic system.

\vskip -0.6cm
\begin{widetext}
\begin{eqnarray}
&\dot{\rho}_{11}=&-i\Omega_{c}(\rho_{12}-\rho_{21})+\gamma_{1} \rho_{22},\nonumber \\
&\dot{\rho}_{33}=&-i\Omega_{e}(\rho_{32}-\rho_{23})+\Gamma_{1} \rho_{22}-\gamma_{2} \rho_{33},\nonumber \\
&\dot{\rho}_{44}=&-i\Omega_{b}(\rho_{42}-\rho_{24})+\Gamma_{2} \rho_{22}-\gamma_{3} \rho_{44},\nonumber \\
&\dot{\rho}_{12}=&-i\Omega_{e}\rho_{13}-i\Omega_{b}\rho_{14}-i\Omega_{c}(\rho_{11}-\rho_{22})-(\frac{\gamma_{1}+\Gamma_{1}+\Gamma_{2}}{2}+i \Delta_{c})\rho_{12},
\nonumber \\
&\dot{\rho}_{13}=&-i\Omega_{e}\rho_{12}+i\Omega_{c}\rho_{23}-(\frac{\gamma_{2}+i\omega_{43}}{2}+i \Delta_{e}+i \Delta_{c}) \rho_{13}\\ \label{eq1}
&\dot{\rho}_{14}=&-i\Omega_{b}\rho_{12}+i\Omega_{c}\rho_{24}-(\frac{\gamma_{3}-i\omega_{43}}{2}+i \Delta_{b}+i \Delta_{c}) \rho_{14}\nonumber \\
&\dot{\rho}_{23}=&-i\Omega_{e}(\rho_{22}-\rho_{33})+i\Omega_{c}\rho_{13}+i\Omega_{b} \rho_{43}-(\frac{\gamma_{1}+\gamma_{2}+\Gamma_{1}+\Gamma_{2}+i\omega_{43}}{2} +i\Delta_{e})\rho_{23},\nonumber\\
&\dot{\rho}_{24}=&-i\Omega_{b}(\rho_{22}-\rho_{44})+i\Omega_{c}\rho_{14}+i\Omega_{e} \rho_{34}-(\frac{\gamma_{1}+\gamma_{2}+\Gamma_{1}+\Gamma_{2}-i\omega_{43}}{2}+i \Delta_{b})\rho_{24},\nonumber\\
&\dot{\rho}_{34}=&+i\Omega_{e}\rho_{24}-i\Omega_{e} \rho_{32}-(\frac{\gamma_{2}+\gamma_{3}}{2}-i\omega_{43})\rho_{34},\nonumber
\end{eqnarray}\end{widetext}
Along with $\rho_{ij}=\rho^{*}_{ji}$ (i,j=1,2,3,4) and $\rho_{11}+\rho_{22}+\rho_{33}+\rho_{44}=1$.
Here $\Omega_{e}$, $\Omega_{b}$ and $\Omega_{c}$ were assumed to be real. $\Delta_{e} = \omega_{p}-\omega_{32}$, $\Delta_{b} = \omega_{p}-\omega_{42}$ and $\Delta_{c} = \omega_{c}-\omega_{12}$ are the detunings of probe and coupling fields, respectively. $\omega_{43}$, $\omega_{32}$, $\omega_{42}$ and $\omega_{12}$ are the
resonant frequencies which associate with the corresponding transitions. $\gamma_{1}$, $\gamma_{2}$ and $\gamma_{3}$ are spontaneous emission rates from the excited state $|2\rangle$ to the ground state $|1\rangle$, from the excited state $|3\rangle$ to the excited state $|2\rangle$ and from the excited state
$|4\rangle$ to the excited state $|2\rangle$, respectively. In Ref.[10,20], the decay rates $\gamma_{1}$, $\gamma_{2}$ and $\gamma_{3}$ are $\gamma_{1}\simeq$ 5.3MHz, $\gamma_{2}\simeq$ 0.67MHz,  $\gamma_{3}\simeq$ 0.67MHz, respectively. In our numerical simulation, we choose these parameters to be units by scaling $\gamma$ = 0.67 MHz, and $\gamma_{1}$=8$\gamma$, $\gamma_{2}=\gamma_{3}$=$\gamma$. Here, the direct transition between the excited states $|3\rangle$ and $|4\rangle$, and the transitions between the excited and ground states  $|3\rangle$, $|4\rangle$ $\leftrightarrow $$|1\rangle$ of the atom are assumed to be forbidden in the dipole approximation. The relaxation rate of coherence among states $|3\rangle$ and  $|4\rangle$ is negligible, these transitions $|3\rangle$$\leftrightarrow$$|1\rangle$ and $|4\rangle$$\leftrightarrow$$|1\rangle$ are non-dipole allowed. The decay rates from states $|3\rangle$ and  $|4\rangle$ to state  $|1\rangle$ are zero, which thus can be safely neglected. When the probe field is weak, i.e., $\Omega_{c}\gg \Omega_{e}, \Omega_{b}$, and the atom is initially prepared in the ground state $|1\rangle$, Equation (1) can be solved in the steady state,
\begin{widetext}\begin{eqnarray}
\rho_{24}=\frac{\Omega_{b}\Omega_{c}\rho^{0}_{21}}{(i \Delta_{b}+i \Delta_{c}-\frac{\gamma_{3}+i \omega_{43}}{2})[i\Delta_{b}-\frac{1}{2}(\gamma_{1}+\gamma_{2}+\Gamma_{1}+\Gamma_{2}+i \omega_{43})]+\Omega^{2}_{c}},
\end{eqnarray}\end{widetext}
\begin{widetext}\begin{eqnarray}
\rho_{23}=\frac{\Omega_{e}\Omega_{c}\rho^{0}_{12}}{(i \Delta_{e}+i \Delta_{c}+\frac{\gamma_{3}+i \omega_{43}}{2})[i \Delta_{e}-\frac{1}{2}(\gamma_{1}+\gamma_{2}+\Gamma_{1}+\Gamma_{2}+i\omega_{43})]+\Omega^{2}_{c}},
\end{eqnarray}\end{widetext}
where
\begin{eqnarray}
\rho^{0}_{21}=\frac{2i \Omega_{c}}{\gamma_{1}+\Gamma_{1}+\Gamma_{2}-2 i \Delta_{c}},\nonumber
\end{eqnarray}
The ensemble electric polarization and magnetization of the atomic medium at the probe field frequency can
be obtained by means of the formula $\vec{P}_{e}(\omega_{e})$ $=$$\epsilon_{0}$$\alpha_{e}(\omega_{ e})$$\vec{E}(\omega_{e})$ and
$\vec{P}_{b}(\omega_{b})$=$\mu_{0}\alpha_{b}\vec{B}(\omega_{b})$, respectively, which are rank 2 tensors and defined by the Fourier transform.
Here, we adopt the explicit $\omega_{e}$ dependence $\alpha_{e}(\omega_{P})\equiv\alpha_{e}$ and set $\vec{E}_{e}$ to parallel to the atomic dipole $\vec{d}_{21}$,
and we set magnetic dipole to be perpendicular to the induced electric dipole in accordance with the classical Maxwell's electromagnetic wave-vector
relation. Then the electric polarizability $\alpha_{e}$ and the magnetization $\alpha_{m}$ are both scalars, and their expressions are as follows\cite{21,22}:

\begin{eqnarray}
\alpha_{e}=\frac{\vec{d}_{32}\rho_{23}}{\epsilon_{0}\vec{E}_{e}}=\frac{\mid
{d_{32}}\mid^{2} \rho_{32}}{\epsilon_{0}\hbar\Omega_{e}},
\alpha_{m}=\frac{\mu_{0}\vec{\mu}_{42}\rho_{24}}{\vec{B}}=\frac{\mu_{0}\mid\mu_{42}\mid^{2}\rho_{24}}{\hbar\Omega_{B}}.\label{eq4}
\end{eqnarray}
In which $\epsilon_{0}$ and $\mu_{0}$ are the permittivity and permeability of vacuum. And the relative permittivity and relative permeability can be given according to the Clausius-Mossotti relations considering the local effect in dense medium\cite{23} as follows:

\begin{eqnarray}
\epsilon_{r}=\frac{1+\frac{2}{3}N\alpha_{e}}{1-\frac{1}{3}N\alpha_{e}},    \mu_{r}=\frac{1+\frac{2}{3}N\gamma_{m}}{1-\frac{1}{3}N\gamma_{m}}.\label{eq5}
\end{eqnarray}
where N is the $^{87}$Rb number density in a vapor cell.

\section{The result and analysis for the simulating $^{87}$Rb atomic system}

In this section, we discuss the left-handedness via the typical optical parameters, i.e., the relative dielectric permittivity $\varepsilon_{r}$ , the relative magnetic permeability  $\mu_{r}$ and refractive index  $ n $ of the cold $^{87}$Rb atomic system. Under the multiply different parameters modulation, the real parts of relative dielectric permittivity $\varepsilon_{r}$, relative magnetic permeability $\mu_{r}$ become negative over the simultaneous frequency band and have the adjustable characteristics, which demonstrates the  adjusting left-handedness in our proposal. As regards their imaginary parts, we pay little attention here.

\begin{figure}[htp]
\centering
\includegraphics[totalheight=1.1 in]{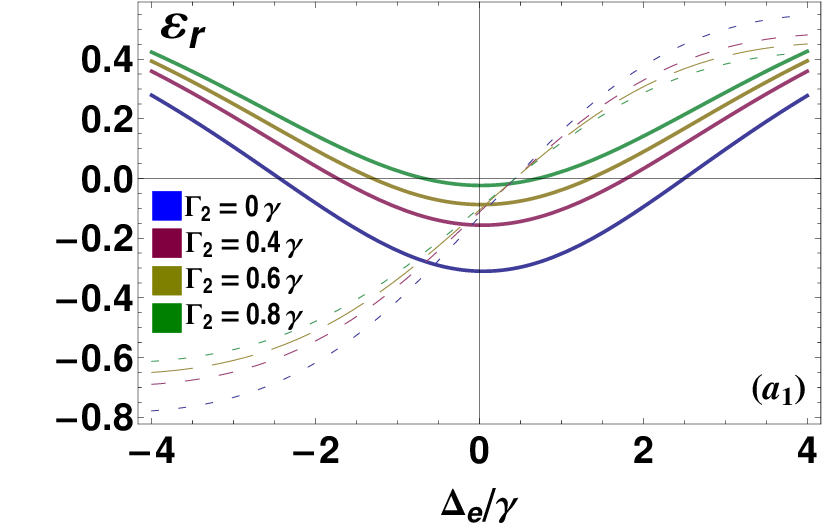}\includegraphics[totalheight=1.1 in]{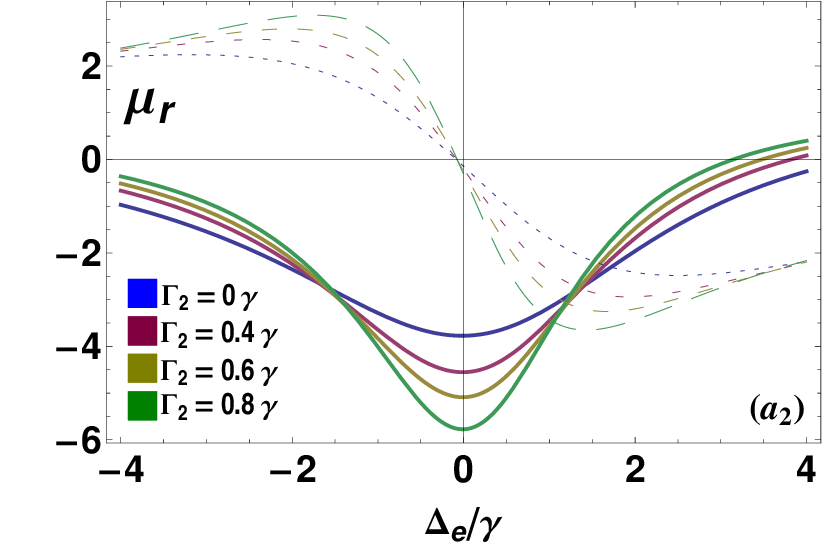}
\includegraphics[totalheight=1.1 in]{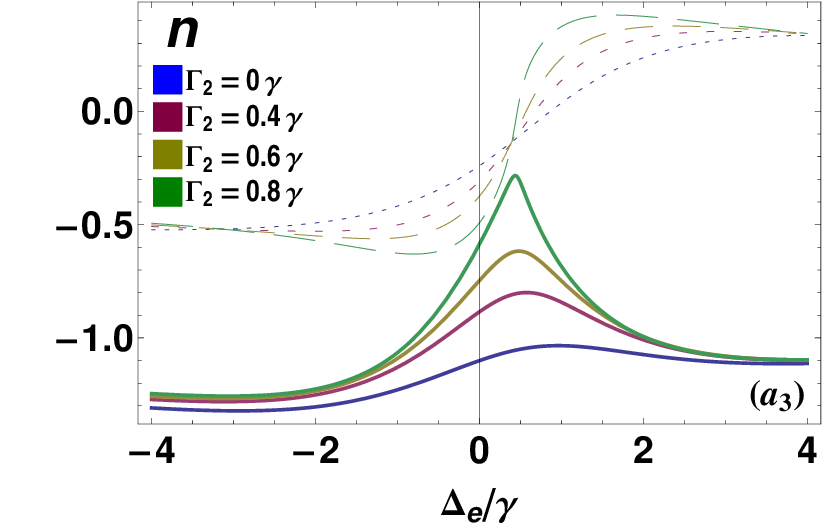}\includegraphics[totalheight=1.1 in]{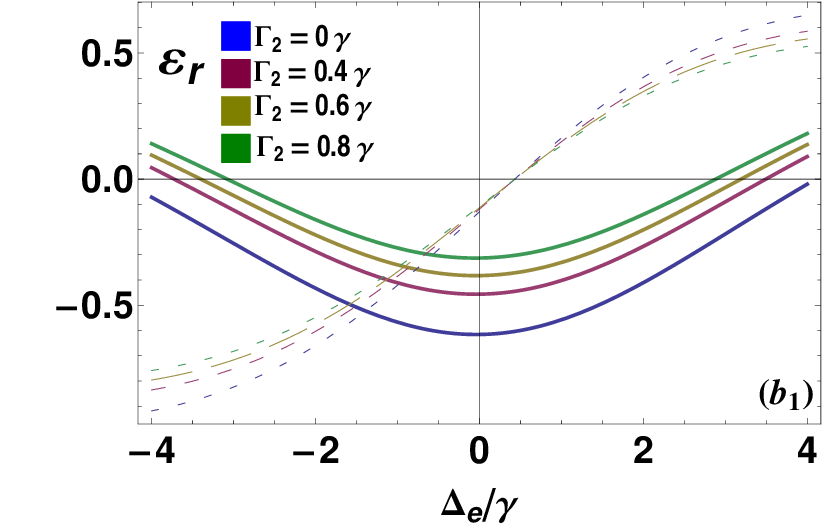}
\includegraphics[totalheight=1.1 in]{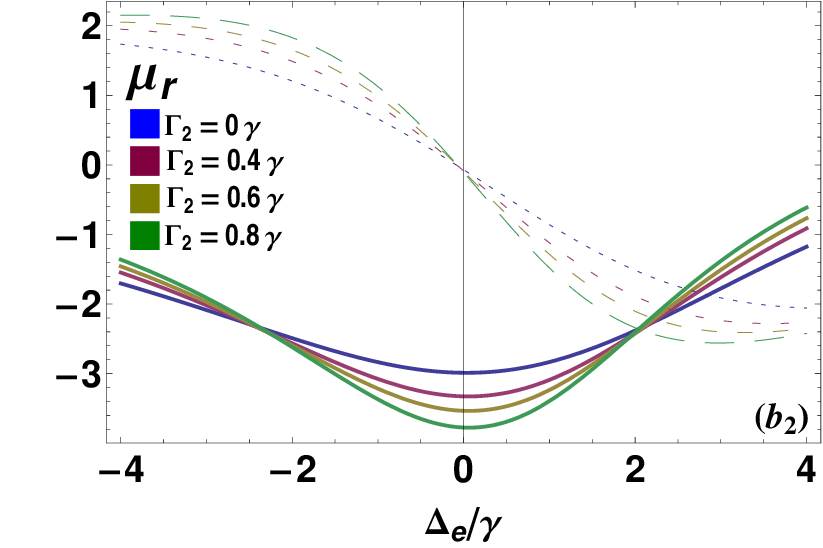}\includegraphics[totalheight=1.1 in]{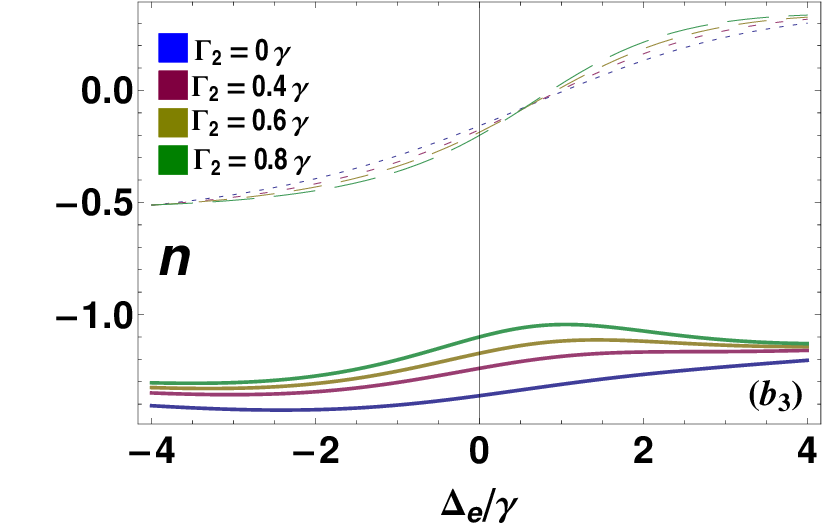}\hspace{0in}%
\caption{(Color online) Real (solid lines) and imaginary (dashed lines) parts of permittivity $\varepsilon_{r}$, permeability  $\mu_{r}$  and refractive index  $ n $ as a function of the rescaled detuning parameter $\Delta_{e}/\gamma$ with $\Gamma_{2}=$0$\gamma$, 0.4$\gamma$, 0.6$\gamma$, 0.8$\gamma$, and $\Omega_{c}=22.5\gamma$, $\Omega_{e}=0.5\gamma$, $\Gamma_{1}$=1.5$\Gamma_{2}$, $\Delta_{c}$=-0.25$\gamma$, $\Delta_{b}$=-1.5$\Delta_{e}$. The $^{87}$Rb number density is chosen as N=$1.04\times10^{21}$ in (a1)$\sim$ (a3) and $1.5 N $ in (b1)$\sim$ (b3).\label{2}}
\end{figure}

First of all, we consider the $^{87}$Rb number density to examine its role in the left-handedness, and the results are presented in Fig. 2 for N=$1.04\times10^{21}$ in (a1) to (a3) and for 1.5N in (b1) to (b3) with four different incoherent pumping frequencies of the two incoherent pumping fields.
In the case of the two incoherent pumping fields are absent, i.e., $\Gamma_{1}$=$\Gamma_{2}$=0, it can be seen from Fig. 2 (a1), the negative frequency band for the real parts of permittivity $\varepsilon_{r}$, i.e., Re[$\varepsilon_{r}$] is the widest, and the frequency ranges for negative Re[$\varepsilon_{r}$] shrink with the increasing of the two incoherent pumping fields. When $\Gamma_{2}=$0.8$\gamma$, the frequency range for negative Re[$\varepsilon_{r}$] comes to the minimum.
The negative real part of permittivity Re[$\varepsilon_{r}$] reaches the maximum at the probe resonant point, i.e., $\Delta_{e}$=0, when the different incoherent pumping fields drive the corresponding transitions. On the contrary, the real part of relative magnetic permeability $\mu_{r}$, i.e., Re[$\mu_{r}$] displays the different characteristic in  Fig. 2 (a2). Especially, at the resonant point, the real part of relative magnetic permeability Re[$\mu_{r}$] increasing with the two enlarging incoherent pumping fields. The two incoherent pumping fields play a destructive role in the negative Re[$\varepsilon_{r}$], while a constructive role in the negative real part of permeability $\mu_{r}$. And the real part of refractive index $ n $ in  Fig. 2 (a3) bounces flexibly near the resonant point when the two incoherent pumping fields pump the corresponding transitions increasingly. In Fig. 2 from (b1) to (b3), the $^{87}$Rb number density is set 1.5 times as those in Fig. 2 from (a1) to (a3). We note that the dense $^{87}$Rb vapor can enlarge the frequency range for negative real parts of permittivity $\varepsilon_{r}$ in  Fig. 2 from (b1) to (b3). In addition to this, the dense $^{87}$Rb vapor results in the decreasing negative response for real parts of permittivity $\varepsilon_{r}$,  permeability $\mu_{r}$ and refractive index  $ n $, respectively, which is also shown in Fig. 2 from (b1) to (b3).

In the following, let us explore the effect of the intensity of the strong coupling field which couples ground level $|1\rangle$ and the intermediate level $|2\rangle$ on the left-handedness in the cold $^{87}$Rb atomic system for four different pumping frequencies of the two incoherent pumping fields. The Rabi frequency $\Omega_{c}$ was set $\Omega_{c}$=28$\gamma$ in Fig. 3 from (c1) to (c3) and $\Omega_{c}$=32$\gamma$ in Fig. 3 from (d1) to (d3), respectively. Comparing Fig. 3 (c1) with Fig. 3 (d1), it's noted that the strong coupling field influences the frequency ranges for negative real parts of permittivity $\varepsilon_{r}$ positively when the two incoherent pumping fields select the same values. The stronger coupling field can enlarge the frequency range for negative Re[$\varepsilon_{r}$], which may flexibly modulate our proposal to obtain negative Re[$\varepsilon_{r}$]. The Re[$\mu_{r}$] in Fig. 3 (c2) and Fig. 3 (d2) show little difference when the diverse strong coupling field couples ground level $|1\rangle$ and the intermediate level $|2\rangle$ besides the frequency ranges for negative values. The Re[$ n $] in Fig. 3 (c3) and Fig. 3 (d3) prove the stronger coupling field can result in the larger negative values with the four different pumping frequencies of the two incoherent pumping fields..

Combining with Re[$\varepsilon_{r}$] and Re[$\mu_{r}$]in Fig. 2 and Fig. 3, we note that the two incoherent pumping field play a destructive role in
Re[$\varepsilon_{r}$], but a constructive role in Re[$\mu_{r}$]. The reason may qualitatively explain as the $^{87}$Rb atom configuration and the atom population.
The incoherent pumping field ($\Gamma_{1}$) drives the transition $|2\rangle$ $\leftrightarrow$$|3\rangle$ with the pumping frequency of 1.5 times as the incoherent pumping field ($\Gamma_{2}$) which drives the transition $|2\rangle$ $\leftrightarrow$$|4\rangle$. The stronger incoherent pumping field ($\Gamma_{1}$) causes more population to $|3\rangle$ which depresses the negative electric response, as is shown by Re[$\varepsilon_{r}$] in Fig. 2 and Fig. 3. The depressed the negative electric response leads to the the destructive Re[$\varepsilon_{r}$]. While the moderately increasing incoherent pumping field ($\Gamma_{2}$) stimulates and causes the constructive Re[$\mu_{r}$] shown in Fig. 2 and Fig. 3.

\begin{figure}[htp]
\center
\includegraphics[totalheight=1.1 in]{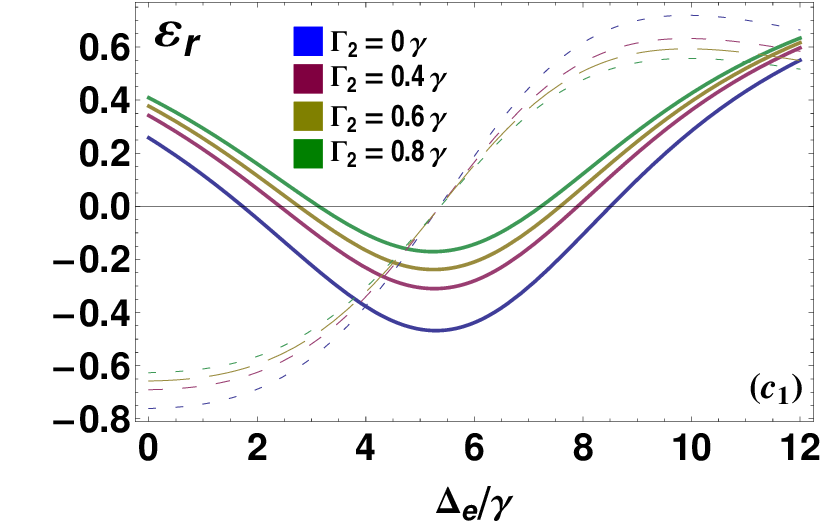}\includegraphics[totalheight=1.1 in]{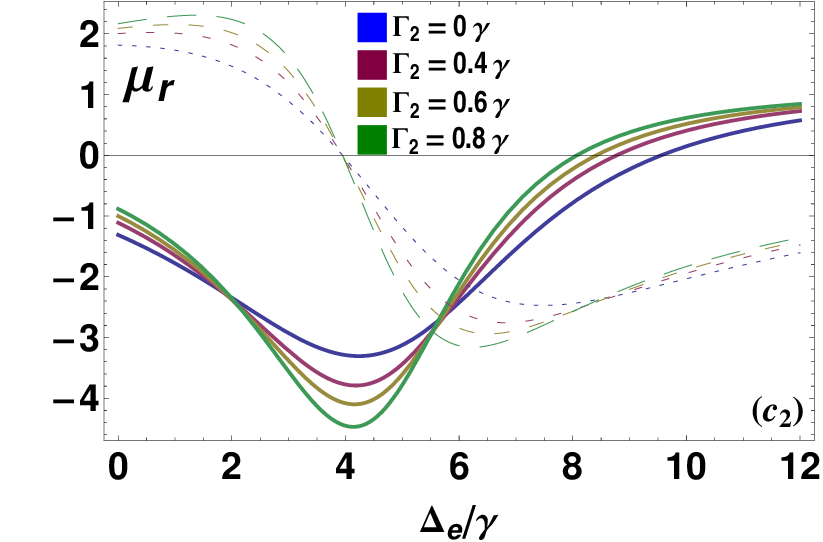}
\includegraphics[totalheight=1.1 in]{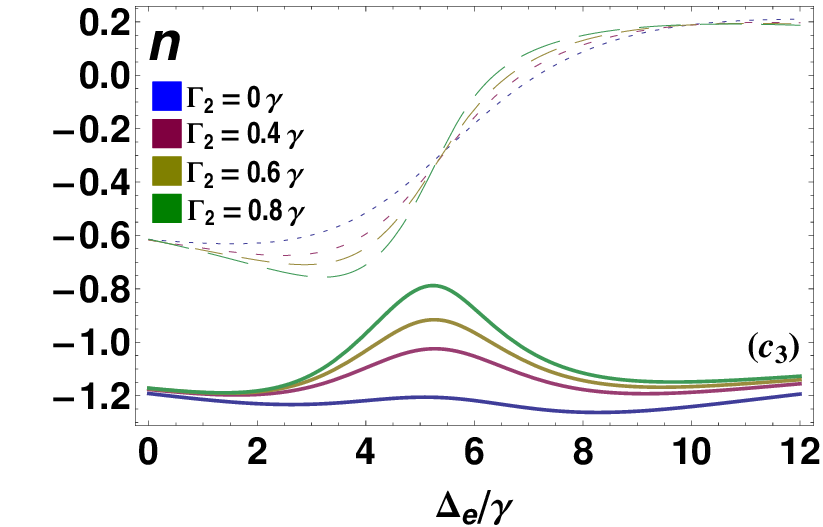}\includegraphics[totalheight=1.1 in]{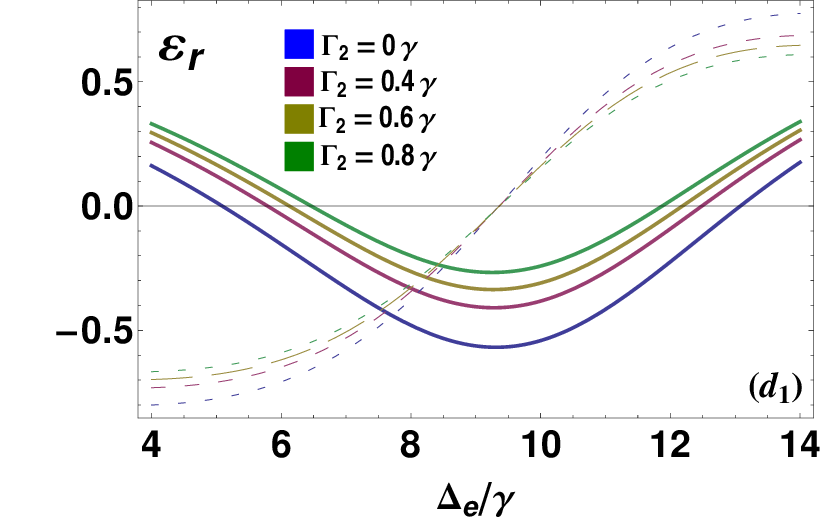}
\includegraphics[totalheight=1.1 in]{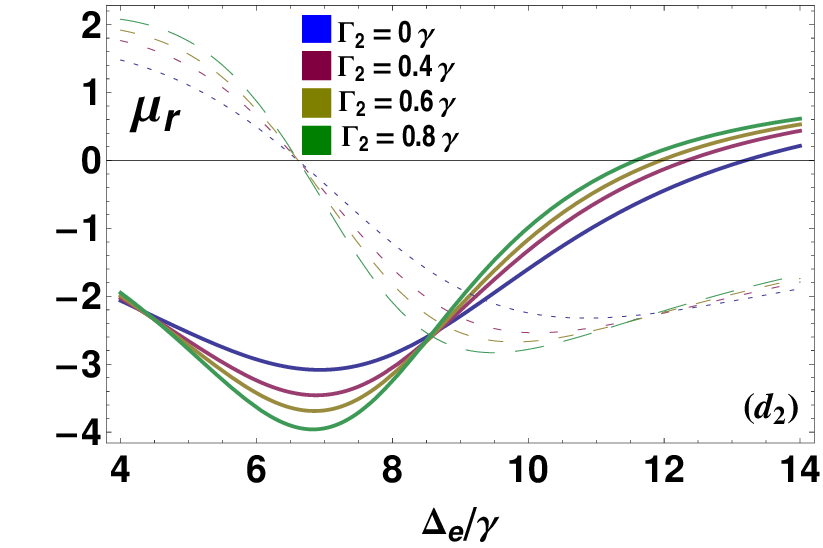}\includegraphics[totalheight=1.1 in]{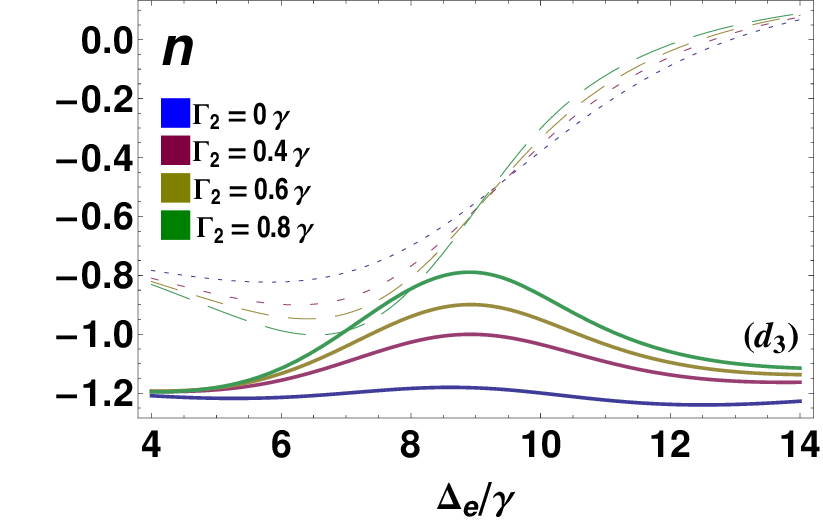}
\caption{(Color online) Real (solid lines) and imaginary (dashed lines) parts of permittivity $\varepsilon_{r}$, permeability  $\mu_{r}$  and refractive index  $ n $ as a function of the rescaled detuning parameter $\Delta_{e}/\gamma$ with $\Delta_{c}$=0.25$\gamma$, the $^{87}$Rb number density $N=1.04\times10^{21}$. The Rabi frequency of the strong coupling field, $\Omega_{c}$=28$\gamma$ in (c1)$\sim$ (c3), $\Omega_{c}$=32$\gamma$ in (d1)$\sim$ (d3) and the other parameters are same as those in Fig.2.\label{3}}
\end{figure}

\section{Conclusions}

In summary, we have investigated the left-handedness in the cold $^{87}$Rb atom by exploring the real parts of three optical parameters, i.e., the relative dielectric permittivity $\varepsilon_{r}$, relative magnetic permeability $\mu_{r}$ and refractive index $ n $. The results clearly show that the increasing $^{87}$Rb number density and stronger coupling field can influence the left-handedness  more greatly, and the increasing incoherent pumping fields depresses the negative electric response while constructs the negative magnetic response. The left-handedness achieved in our proposal is based on the practical $^{87}$Rb atomic hyperfine structure which may increase the probability for realization. In addition, the left-handedness adjusted by multiple parameter provides the flexibility and possibility for the coming experiment.

\end{document}